\documentclass[10pt]{article}
\usepackage[T1]{fontenc}
\usepackage{times}
\usepackage{authblk}
\usepackage{amssymb,amsmath}
\usepackage[letterpaper,margin=1in]{geometry}
\usepackage{graphicx}
\usepackage{natbib}
\usepackage{enumitem}
\usepackage{capt-of}
\usepackage[compact]{titlesec}
\usepackage{setspace}

\title{Earth and Terrestrial Planet Formation}
\date{}
\author[1,2]{Seth A. Jacobson\thanks{seth.jacobson@oca.eu}}
\author[3]{Kevin J. Walsh}
\affil[1]{Bayerisches Geoinstitut, Universt{\"a}t Bayreuth, 95440 Bayreuth, Germany}
\affil[2]{Laboratoire Lagrange, Observatoire de la C{\^o}te d'Azur, 06304 Nice, France}
\affil[3]{Planetary Science Directorate, Southwest Research Institute, 80302 Boulder, CO, USA}

\begin{document}
\maketitle
\section*{Abstract}
The growth and composition of Earth is a direct consequence of planet formation throughout the Solar System. We discuss the known history of the Solar System, the proposed stages of growth and how the early stages of planet formation may be dominated by pebble growth processes. Pebbles are small bodies whose strong interactions with the nebula gas lead to remarkable new accretion mechanisms for the formation of planetesimals and the growth of planetary embryos.

Many of the popular models for the later stages of planet formation are presented. The classical models with the giant planets on fixed orbits are not consistent with the known history of the Solar System, fail to create a high Earth/Mars mass ratio, and, in many cases, are also internally inconsistent. The successful Grand Tack model creates a small Mars, a wet Earth, a realistic asteroid belt and the mass-orbit structure of the terrestrial planets.

In the Grand Tack scenario, growth curves for Earth most closely match a Weibull model. The feeding zones, which determine the compositions of Earth and Venus follow a particular pattern determined by Jupiter, while the feeding zones of Mars and Theia, the last giant impactor on Earth, appear to randomly sample the terrestrial disk. The late accreted mass samples the disk nearly evenly.
\section{Introduction}
The formation and history of Earth cannot be studied in isolation. Its existence, composition and dynamics are a direct consequence of events that occurred throughout the Solar System. Planet formation is a combination of local and global processes at every stage of growth. Therefore, the early Earth must be placed in the context of the history of the Solar System. 
\subsection{History of the Solar System}
\label{sec:history}
There are very few firmly established temporal events during planet formation, however one that seems irrefutable is that the giant planets must have accumulated all of their gas envelopes, nearly all of their mass, by the time the nebular gas is removed from the disk. Observations of nebular disks about other stars indicate that disk survival times are typically only millions of years~\citep{Haisch:2001bx,Briceno:2001fs,Mamajek:2009de}, although there is some evidence that disks can last for tens of millions of years~\citep{Pfalzner:2014hp}. While we do not have a direct observation of our own disk survival time, we do know that chondrule formation--a likely nebular process~\citep{Alexander:2008ih} ended after a few million years~\citep{Kita:2005ws,Villeneuve:2009ji}. Furthermore, there is evidence from modeling of the formation of Iapetus including $^{26}$Al decay that it mush have formed $\sim$3.4 to 5.4 My after CAIs~\citep{CastilloRogez:2007gx,CastilloRogez:2009kh}. Since the formation of Iapetus requires Saturn to be near completion, this also constrains the age of Saturn.

The last giant impact on Earth, the Moon-forming impact, has a date established to be approximately 70--110 My after CAIs from three lines of evidence: radiometric~\citep{Touboul:2007is,Allegre:2008iq,Halliday:2008bo}, dynamical-elemental~\citep{Jacobson:2014cm} and dynamical-isotopical~\citep{Bottke:2014uu}. There is some disagreement, others have suggested earlier dates using different models to interpret radiometric data~\citep{Yin:2002ex,Jacobsen:2005bj,Taylor:2009jr}, but even these 20--60 My estimates are significantly after the few million year lifetime of the nebular gas disk. Interestingly, the age of Mars is considerably different than that of Earth. Radiometric evidence from the Hf-W system indicates that it likely formed within 10 My~\citep{Nimmo:2007bg} and perhaps it was within a few percent of its final mass by the time the disk dissipated~\citep{Dauphas:2011ec}, so Mars may have nearly completed its formation in the presence of the nebular gas. The formation dates of Mercury and Venus are unknown.

The other temporal event with significant evidence in the history of the Solar System is the late heavy bombardment. This intense period of impacts $\sim$400 My after CAIs left a record on the Moon~\citep{Tera:1974jk,Ryder:1990ip,Cohen:2000bb,Ryder:2002gd}, on Earth~\citep{Marchi:2014gm}, and on Vesta through analysis of the HED meteorites~\citep{Marchi:2013du}. From the crater record on the Moon, it appears that $\sim$10$^{-4}$ M$_\oplus$ of material was delivered to Earth during this bombardment~\citep{Morbidelli:2012ko,Marchi:2014gm}. For a contrasting interpretation of the lunar cratering record and asteroid melt chronology, see~\citet{Chapman:2007bt}, although this review occurred before the terrestrial and HED evidence was presented.
\paragraph{Nice model} 
\label{sec:Nice}
There is significant evidence in the orbital structure of the Kuiper Belt that the giant planets migrated due to interactions with this disk of outer planetesimals~\citep{Fernandez:1984iy,Malhotra:1995dy,Gomes:2003cd,Levison:2003wq}. This interaction caused the outward migration of Saturn, Uranus and Nepture and the inward migration of Jupiter because the planetesimals are first inwardly scattered by the outer three giant planets and then scattered out of the Solar System by Jupiter. When this migration occurred is not directly known, but it has been associated with the origin of the late heavy bombardment through a dynamical instability~\citep{Strom:2005ur}; the `Nice' model stipulates that an early compact configuration of the giant planets undergoes an instability~\citep{Gomes:2005gi}. This giant planet instability explains a number of features of the Solar System including the Trojans of Jupiter~\citep{Morbidelli:2005dr} and Neptune~\citep{Tsiganis:2005dg}, the number and inclination distribution of the irregular satellites of the giant planets~\citep{Nesvorny:2007dv,BottkeJr:2010gb}, and dynamical features of the Kuiper Belt~\citep{Levison:2008ib,Morbidelli:2014jw}. Furthermore, close encounters between an ice giant and Jupiter reproduce the structure of the main asteroid belt~\citep{Morbidelli:2010kj}. More recently, the model has been updated, Nice 2.0, to better reflect the likely orbital spacing of the giant planets in mean motion resonances after removal of the gas~\citep{Levison:2011gt}.

While a complete rejection of the Nice model would require a number of less likely or so-far incomplete hypotheses to take its place, the connection between the late heavy bombardment and the giant planet instability of the Nice model is not required. For instance, a Nice model-like instability may occur immediately after nebula gas removal, thereby creating the Solar System architecture and small body populations that are observed today. In this alternative scenario, the late heavy bombardment would either need to be triggered by the instability of a small fifth planet at $\sim$2 AU~\citep{Chambers:2007cu}, to not occur due to a different interpretation of the absolute age evidence on the Moon~\citep{Chapman:2007bt}, or to be triggered by the catastrophic break-up of a Vesta-sized Mars-crossing asteroid due to a different interpretation of the cratering evidence on the Moon~\citep{Cuk:2010hw,Cuk:2012bg}. The latter two alternatives are unlikely because of new evidence that the late heavy bombardment is recorded in the HED meteorites~\citep{Marchi:2013du}.

It is important to note that the Nice model and the Grand Tack model are not the same and are independent of one another. The inward-then-outward migration of Jupiter and Saturn called the `Grand Tack'~\citep{Walsh:2011co}, which is described in Section~\ref{sec:grandtack}, is proposed to take place in the presence of the nebular gas, so much earlier in the history of the Solar System than the Nice model. The Grand Tack scenario is consistent with alternatives to the Nice model such as the conjecture that the giant planet instability occurs shortly after gas removal and the evidence for the late heavy bombardment has some other explanation such as the Planet V hypothesis~\citep{Chambers:2007cu}.
\subsection{Stages of planet formation}
For convenience, we divide the history of planet formation into four stages: dust sedimentation and growth, planetesimal growth, planetary embryo growth, and planet growth. While it may be unlikely that these stages are coincident locally, it's very likely that they are occurring simultaneously at different radii across the disk since the orbital period, solid surface density, and gas density, viscosity and temperature change dramatically with semi-major axis as well as time. Particularly, there is a divide in the Solar System between the giant planets, Jupiter and Saturn, which grew large enough to accrete most of their mass from the solar nebula, doing so during the first few million years of Solar System history before the nebular gas dissipated, and the terrestrial planets, which did not grow large enough to accrete a significant amount of solar nebula gas. Of course, the sub-giants Uranus and Neptune grew at a rate somewhere in between, but exterior to Jupiter and Saturn.

It is computationally difficult (and impossible as of now) for a single numerical simulation to cover all stages of growth, so typically each stage is handled separately. Earlier processes such as dust growth and planetesimal formation are often studied in local patches of the disk or in small annuli. This is thought to be acceptable since perturbations from elsewhere in the disk are homogenous across the studied zone. Even global simulations of the later stages are computationally expensive. 

Each growth stage is marked by a different dominant accretion mechanism for the largest bodies---those we typically care the most about, but smaller bodies may continue to accrete according to earlier paradigms. We review each growth stage, in turn, focusing on new developments, but we begin by defining some new terminology that has appeared in the planet formation literature.

The motion of solid objects in a gas disk is dictated by the Stokes number. It is a dimensionless number with a long history in fluid mechanics that quantifies the momentum coupling between an object of approximate radius $R$ and the surrounding fluid. In the context of a gas disk, it is defined: $St \approx (\rho_s / \rho_0) R \Omega / c_0$ where $\rho_s / \rho_0$ is the ratio of the solid to mid-plane gas density, $\Omega$ is the Keplerian orbital frequency, and $c_0$ is the mid-plane sound speed. This formula is correct for Epstein drag regime, which assumes that the gas mean free path is longer than the object~\citep[for a review of other drag regimes and the appropriate formulae for determining the Stokes number, see][]{Youdin:2010ji}. Accretion depends both on the Stokes number of the target and projectile~\citep[for recent examples see][]{Chambers:2014gg,Guillot:2014te}. When $St \ll 1$, the object is completely entrained in the gas and when $St \gg 1$, the object's motion is independent of the presence of the gas. Thus different regimes are identified: objects with Stokes numbers much less than 0.01 are `dust', those with Stokes numbers between 0.01 and 1 are `pebbles', and those with Stokes numbers just greater than 1 are `boulders'. Planetesimals and embryos have Stokes numbers much greater than 1. From the definition of the Stokes number, an object's membership in each regime is dependent on the disk properties. The Stokes number is the fundamental quantity determining the significance of passive concentration processes, streaming instabilities, and pebble accretion; all discussed in detail in Section~\ref{sec:pebbles}.
\paragraph{Dust sedimentation and growth}
Planet formation begins in the nebular disk that forms simultaneously with the protostar from a collapsing molecular cloud as a result of the conservation of angular momentum. Pre-existing dust grains as well as condensates from the gas sediment towards the mid-plane of the protoplanetary disk~\citep{Weidenschilling:1980cq}. Turbulent diffusion of solid material due to entrainment with the gas eventually balances gravity, and so a vertical equilibrium structure is assembled in the disk~\citep{Weidenschilling:1993ws,Cuzzi:1993co,Dubrulle:1995jn,Carballido:2006ej}. In this structure, dust grains can grow via binary collisions accreting through adhesive forces, `fluffy' aggregation and compaction, and by mass transfer during fragmenting collisions~\citep[for reviews, see][]{Dominik:2007wg,Johansen:2014vw}. From these growth mechanisms, a population of pebbles and boulders emerges from the coagulating dust.
\paragraph{Planetesimal growth}
All objects experience some orbital drag because the nebular gas is partially pressure supported; a parcel of gas orbits the Sun at sub-Keplerian speeds but boulders orbit at Keplerian speeds. Thus boulders experience a headwind, which generates energy loss and their orbits spiral into the Sun~\citep{Whipple:1972vv}. This effect is often referred to as the `radial drift' or `meter' barrier, when considering specific disk conditions. Furthermore, as objects grow aerodynamic drag decreases, impact velocities increase, and binary collisions are no longer an efficient growth mechanism because colliding boulders bounce or even fragment~\citep[for review, see][]{Blum:2008fi}. This is often referred to as the `bouncing' barrier~\citep{Zsom:2010hg}.

These boulder barriers restrict growth to $St \sim 1$ or about $\sim0.1$--$1$ m at an AU for a standard model disk~\citep{Weidenschilling:1977wt,Zsom:2010hg}. It is possible that fluffy aggregates of dust grains can grow to km or even larger sizes if ice enhances stickiness and collision velocities remain below $\sim$50 m s$^{-1}$~\citep{Wada:2009dv,Wada:2013do,Okuzumi:2012kd}; whether these planetesimals would be too water rich to contribute significantly to terrestrial planet formation is an open question. In Section~\ref{sec:pebbles}, we discuss new ideas regarding the interactions between gas and `pebbles' that would create $\sim100$ km planetesimals directly from $<$1 m `pebbles' skipping over these barriers~\citep{Johansen:2007cl,Cuzzi:2008js}.
\paragraph{Planetary embryo growth}
Objects that grow beyond the boulder barriers are planetesimals, and they grow at different rates depending on their size. Smaller planetesimals in the `orderly' growth regime accrete according to their physical cross-sections alone and so growth proceeds slowly~\citep{Safronov:1972tr,Nakagawa:1983ij}. It's possible that most large planetesimals including those that continued to grow into planetary embryos and planets bypassed much of this growth regime and grew suddenly through gravo-turbulent growth mechanisms from pebble and boulder sizes to hundreds or thousands of kilometers across (a description of this growth and its consequences is in Section~\ref{sec:pebbles}).

If planetesimals have grown large enough that the escape velocities from their surfaces match or exceed their relative velocities, then gravitational focusing is important~\citep{Safronov:1969cg,Greenberg:1978ea}. The enhancement of the collision cross-section due to gravitational focusing is proportional to the square of the ratio of escape to relative velocities~\citep[for a derivation and further review;][]{Armitage:2014wk}. Since dynamical friction reduces the relative velocities of the largest bodies compared to the rest and those bodies already have the highest escape velocities, planetesimal growth transforms into a `runaway' growth process; the largest bodies have the largest relative binary collision cross-sections and hence grow the fastest and become planetary embryos~\citep{Wetherill:1989ev,Wetherill:1993fp,Ida:1992el,Kokubo:1996be}. While gas drag is less efficient for larger bodies, dynamical friction efficiently transfers energy from these larger bodies to smaller bodies that interact with the gas more strongly~\citep{Wetherill:1993fp,Kokubo:1996be}. Dynamical friction is the damping of the eccentricities and inclinations of larger bodies due to gravitational interactions with a large number of smaller bodies. Smaller bodies are also more likely to fragment rather than grow due to these higher relative velocities and their lower gravitational binding energies, further increasing the number of even smaller bodies and the effectiveness of dynamical friction and gas drag~\citep{Wetherill:1989ev}. 

Runaway growth produces a bi-modal mass distribution in the disk of planetary embryos and planetesimals. The formation of planetary embryos is self-limiting and slows down runaway growth by viscously stirring the nearby planetesimal population increasing relative velocities~\citep{Lissauer:1987ig,Ida:1993ch}, so growth is slower and similar between neighboring embryos. This `oligarchic' growth maintains the bi-modal mass distribution of planetary embryos so individually each embryo is about 100 times as massive as the average planetesimal.  Orbital repulsion keeps planetary embryos about 10 mutual Hill radii apart from each other~\citep{Kokubo:1995ib,Kokubo:1998ka}. See~\citet{Morbidelli:2012iz} and~\citet{Raymond:2013wy} for more mathematical reviews of the runaway and oligarchic dynamical processes.
\paragraph{Planet growth, the giant impact phase}
Oligarchic growth ends when energy transfer via dynamical friction can no longer continue circularizing the orbits of the planetary embryos. Typically, this occurs because the planetesimal population becomes too depleted as planetesimals are accreted onto planets and the Sun or scattered from the Solar System. Many simulations do not include the role of collisional grinding but this also leads to a reduction of the planetesimal population~\citep{Levison:2012wp}. If gas drag was a major source of dissipation, then the sudden removal of the solar nebula can trigger the end of the oligarchic regime~\citep{Iwasaki:2002kh,Zhou:2007ex}. 

Without dynamical friction, planetary embryos perturb each other onto crossing orbits leading to either giant impacts or scattering events~\citep{Wetherill:1985by}. The planetesimal population steadily declines during this phase until only a few planetesimals are left on semi-stable orbits in the Main Belt. These giant impacts are the last phase of planet formation. Once the remaining planetary embryos have found stable orbits and finished accreting, we call them planets. An extended giant impact phase is consistent with the late formation of Earth~\citep[$\sim$70--110 My;][]{Touboul:2007is,Allegre:2008iq,Halliday:2008bo,Jacobson:2014cm,Bottke:2014uu}. The early formation of Mars~\citep[$\sim$1--3 My;][]{Nimmo:2007bg,Dauphas:2011ec} may imply that it finished its growth during the oligarchic growth phase, so it is a stranded embryo that never participated in the giant impact phase~\citep{Jacobson:2014it}. 

In the outer Solar System, the embryos need to grow to $\sim$10 M$_\oplus$ giant planet cores, so that gas accretion occurs~\citep{Pollack:1996jp}. The core accretion scenario of dust growth, runaway growth and oligarchic growth is too slow~\citep[for a review of the difficulties in this scenario, see][]{Levison:2010bg}. Recently, an alternative solution has been proposed called pebble accretion~\citep{Ormel:2010ii,Johansen:2010df,MurrayClay:2011vw,Lambrechts:2012gr,Chambers:2014gg}. While this pebble growth process was developed to solve a problem in the outer Solar System, it has strong implications for terrestrial planet formation.
\subsection{Pebble growth processes}
\label{sec:pebbles}
In the terrestrial protoplanetary disk, radial drift and the bouncing barrier frustrate growth at boulder sizes preventing planetesimal formation. These barriers may be overcome by what may broadly referred to as gravo-turbulent growth~\citep[for a thorough review, see][]{Johansen:2014vw}, a possibly misleading nomenclature since turbulence isn't always necessary. Gravo-turbulent mechanisms bring together such great particle concentrations that the collection of small particles collapses under self-gravity. By gathering enough dust or pebbles, a single body forms, which is significantly larger than either of the barriers, skipping the boulder size regime altogether. The initial asteroid size-frequency distribution of the Main Belt may have been very shallow and dominated by asteroids larger than 100 km~\citep{Morbidelli:2009dd}, but this is not conclusive since growth from km sized asteroids can also match the current size-frequency distribution of the Main Belt~\citep{Weidenschilling:2011kt}. However, there is also evidence from the Kuiper Belt that large equal mass binary systems with similar surface appearances were likely formed from a gravitationally collapsing cloud of particles that has too much angular momentum for a single body~\citep{Nesvorny:2010da}. These  clues suggest that planetesimals appeared as 100--1000 km bodies.

The importance of gas-particle interactions have a long history~\citep[e.g.][]{Goldreich:1973cr,Weidenschilling:1993ws}, but older ideas focused on creating a very thin, dynamically cold mid-plane layer of planetesimals. This disk could undergo a gravitational instability, but turbulent coupling with the gas and dynamical self-stirring makes this process difficult~\citep{Weidenschilling:1993ws,Youdin:2002iz}. The more recently developed pebble mechanisms that allow growth beyond the boulder barriers fall into two categories. First, gas dynamics drive particles together via `passive' concentration mechanisms. In regions of the disk with significant turbulence, eddies form at the smallest turbulent scales and these drive dust into the regions between the eddies~\citep{Cuzzi:2008js,Chambers:2010ct}. This mechanism creates smaller planetesimals, which may not be consistent with the evidence above. Larger vortices or radial velocity variations created by gap-clearing giant planets, phase transitions in the disk chemistry, or ionization levels of the disk, so called `dead zones' create pressure bumps that collect particles~\citep[for a list of references see section 4.1.4 in][]{Johansen:2014vw}. A pressure bump creates a zonal flow which slows and possibly reverses the radial drift due to gas drag of pebbles~\citep{Whipple:1972vv}. As mentioned, the pressure bumps are associated with specific structures in the disk, so if this mechanism is dominant, growth beyond the boulder barriers only occurs at unique locations. How numerous these locations are and where they are in the disk are open questions, but the answers so far appear to be few and typically they rest outside of the terrestrial planet formation region~\citep{Johansen:2014vw}.

The second concentration mechanism is not associated with a pre-existing disk structure, instead the `streaming instability' is a consequence of pebble and gas interactions. A single pebble causes almost no back-reaction on the gas dynamics, but a small clustering of pebbles causes the local gas to move with the pebbles at Keplerian orbital rates. Exterior pebbles catch up with this cluster via radial drift and increase the size of the back-reaction, further retarding the radial drift of the cluster and increasing the number of pebbles that catch up with the cluster~\citep{Johansen:2007cl}. Eventually the cluster grows large enough to collapse due to self-gravity creating 100 to 1000 km directly from pebbles~\citep[for a thorough review of the material in the last two paragraphs, see][]{Johansen:2014vw}. 

Pebble-pile planetesimals are an open topic of research~\citep{Hopkins:2014vt,Jansson:2014wi}, but if this mechanism is important there are exciting repercussions. For instance, a fundamental unit of the Solar System is chondrule. Chondrules are small melt spherules found in the most abundant meteorite classes. If chondrules are the pebbles in these processes that would wonderfully explain their ubiquity, but currently most pebble concentration models require particles to be about an order or two of magnitude larger than a typical chondrule given standard disk models~\citep{Johansen:2007cl}. Another repercussion of pebble concentration models are that the variations between planetesimal compositions may be a consequence of where in the disk it is most likely for the streaming instability to occur at a particular time. Due to radial drift, pebbles are constantly migrating from the exterior of the disk inwards. If there is an initial compositional gradient or a temporal gradient associated with dust sedimentation, then this gradient may be frozen in as a function of streaming instability location. Alternatively, pebbles may be the fragments from collisions between larger bodies~\citep{Kobayashi:2010fa,Chambers:2014gg} or fluffy aggregates created from icy dust grains~\citep{Wada:2008eh,Okuzumi:2012kd}. It is not clear what the dominant pebble creation process is or, even, their appearances.

Pebbles may also be important for the rapid growth of planetary embryos~\citep{Ormel:2010ii,Johansen:2010df,MurrayClay:2011vw,Lambrechts:2012gr,Chambers:2014gg}. Pebbles are not entrained along streamlines in the nebular gas but neither are their motions independent of the gas. This has consequences for embryo accretion. As described above, planetesimals are accreted by embryos according to their mutual gravitational cross-sections, however this does not account for gas-solid interactions in the presence of a growing embryo. Pebbles can be accreted from a much larger cross-section than the gravitational focusing cross-section because as they react to the gravitational acceleration from the embryo, they experience gas drag, which reduces their velocity relative to the embryo and this increases the time at which the embryo's gravity acts on them, and so on. 

In the outer Solar System, this process may allow the formation of the giant cores early enough to accrete gas envelopes~\citep{Lambrechts:2014iq,Lambrechts:2014jy}, although the cores could set-up a different persistent oligarchic growth regime~\citep{Kretke:2014wz}. In the inner Solar System, it also could play a significant role~\citep{Levison:2014uj}. If pebbles are a significant portion of the accreted mass, then embryos could become layered with pebbles. If these pebbles are chondrules or clusters of chondrules, then these layers may be an explanation for the remnant magnetic fields and other evidence that certain chondrites were at the surface of larger bodies~\citep[this also assumes that these growing embryos have internally generated magnetic fields;][]{Weiss:2010bz,Carporzen:2011bp,ElkinsTanton:2011fk}. Then again, these larger bodies could have formed directly from pebbles via a passive concentration or streaming instability mechanism. Most radically, if the streaming instability is inefficient, then, perhaps, the five largest planetesimals grew via pebble accretion into the four terrestrial planets and Theia, the Moon-forming impactor---these are the only planet-sized bodies in the inner Solar System for which we have direct evidence of their existence.

The consequences for pebble processes are revolutionizing planetesimal formation and the giant planets, but their repercussions for the terrestrial planets have not been fully explored.
\subsection{New pebble model from dust to embryo}
Assuming that pebble processes are the key to bypassing the boulder barriers, modified planet formation timeline stages emerge. Pre-existing grains or condensates settle into a dust structure centered on the mid-plane, and grow to pebble and boulder sizes from binary collisions. However, some as-yet-unknown fraction of pebbles are concentrated by the streaming instability and collapse directly into large 100---1000 km planetesimals. If this process is successful, then runaway growth occurs from this sea of planetesimals and planetary embryos emerge as oligarchs. If the population of planetesimals remains sparse, i.e. the streaming instability is rare, then embryo formation relies upon pebble accretion. Since the efficiency of pebble accretion is a sharp function of the size of the protoplanet, then planetesimals can be stranded at small sizes never experiencing rapid pebble accretion before the gas in the disk dissipates~\citep{Lambrechts:2012gr}. Interestingly, a bi-modal mass distribution between planetesimals and embryos seems inevitable regardless of the growth paradigm. In either paradigm, though, it's unclear what the distribution of embryo masses are and how much mass is in the embryo population compared to the planetesimal population.

This new model regarding the early stages of planet formation has strong implications for how possible initial compositional gradients are emplaced in the disk, for the timing of different accretion events relative to the decay of radioactive elements, etc. Thus, it requires much future study, but rather than waiting until these early stages are completely understood, studies of terrestrial planet formation have plunged ahead experimenting with different assumed initial distributions of planetesimals and embryos. 
\section{Models of the giant impact phase of terrestrial planet formation}
\begin{figure}[t!]
\begin{center}
\includegraphics[width=\textwidth]{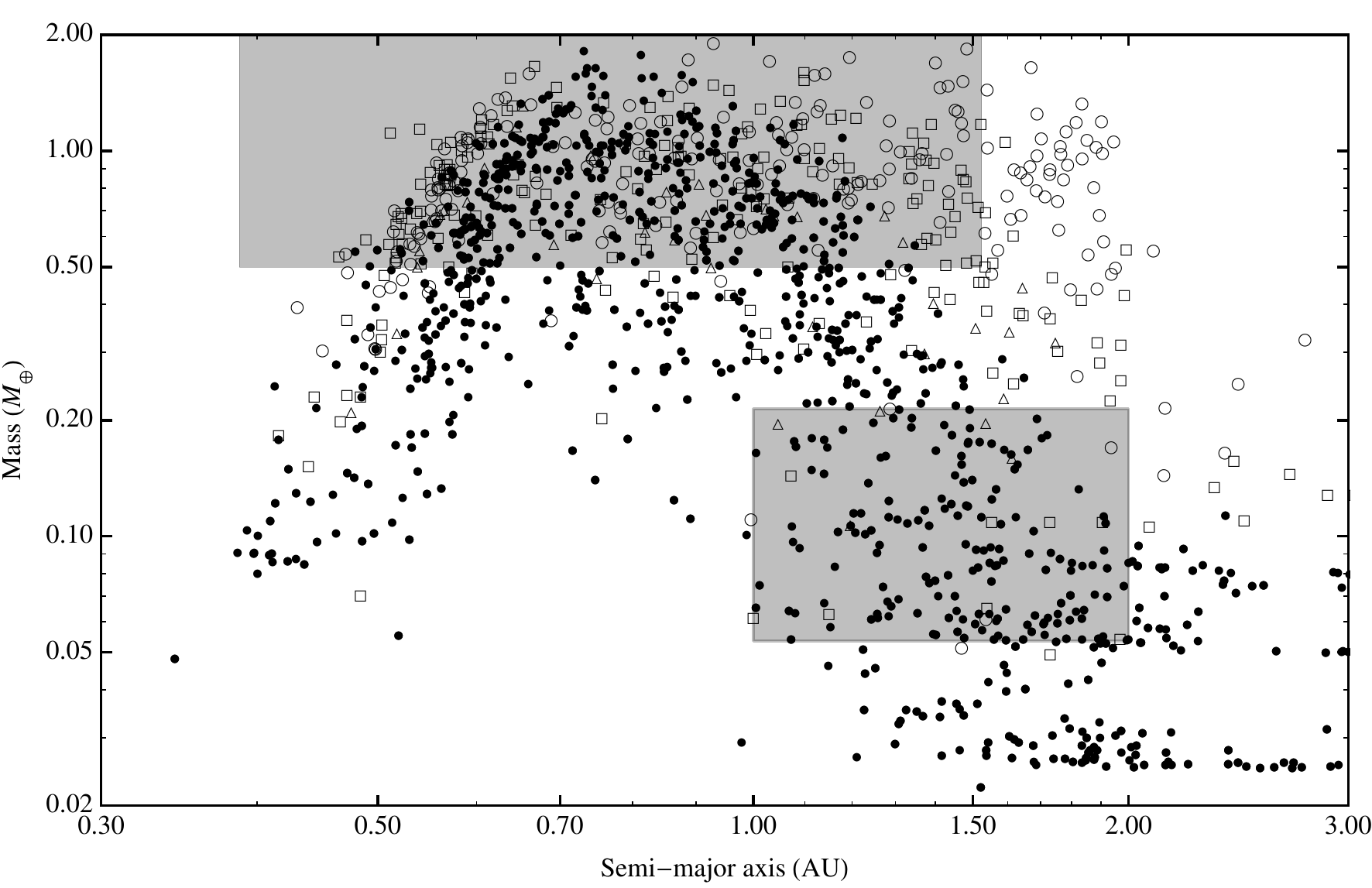}
\caption{The mass and semi-major axis distribution of planets formed from the eccentric~\citep[empty squares;][]{OBrien:2006jx,Raymond:2009is,Fischer:2014fm}, circular~\citep[empty circles;][]{OBrien:2006jx,Raymond:2009is,Fischer:2014fm}, and extra-eccentric~\citep[empty triangles;][]{Raymond:2009is} Jupiter and Saturn models and the Grand Tack model~\citep[solid points;][]{Walsh:2011co,OBrien:2014bk,Jacobson:2014cm,Jacobson:2014it}. Planets within the upper gray rectangle are considered Earth-like and those within the lower gray rectangle are considered Mars-like.}
\label{fig:massplot}
\end{center}
\end{figure}
Unlike earlier stages, which are often studied in local patches of the disk or in small annuli, the giant impact stage of terrestrial planet formation is usually modeled as a nearly global simulation. It is `nearly global' because most N-body simulations simplify the treatment of the giant planets. They are often assumed to be fully formed, because they needed to accrete their gaseous envelopes during the lifetime of the gas disk, and often Uranus and Neptune are neglected since their influence on the terrestrial disk is weak~\citep[e.g. this was found to be so for the Grand Tack;][]{Walsh:2011co}. 

The terrestrial protoplanetary disk at this time is assumed to have a bi-modal mass distribution containing planetary embryos and planetesimals. Such a distribution is produced by both  the ordered, runaway, and oligarchic growth scenario and a pebble-dominated scenario, and it is not clear which scenario is correct. The key parameter is the ratio of total mass in the embryo population to that of the total mass in the planetesimal population. It determines the amount of dynamical friction in the disk. Assuming ordered planetesimal growth, N-body simulations of the planetary embryo growth stage predicted a ratio near unity~\citep{Kokubo:1998ka}, but planetesimal grinding may significantly enhance this ratio over time~\citep{Levison:2012wp} or perhaps, the planetesimals were only ever very small and collisional grinding and radial drift leaves behind a population of only Mars-sized embryos~\citep{Kobayashi:2013bc}. For pebble accretion,  this ratio is unknown. Modern numerical N-body simulations have experimented with a large number of ratio choices~\citep{Jacobson:2014it} or self-consistently model all the stages of planet formation~\citep{Walsh:2014uv}, and they are concluding that a high total embryo mass relative to the total planetesimal mass best reproduces the late accretion record on Earth~\citep{Jacobson:2014cm}.

It is the orbits of Jupiter and Saturn that primarily distinguish the different models of the giant impact phase of terrestrial planet formation. The dynamical excitation of the giant planets is directly reflected in the dynamics of the terrestrial disk via secular resonances (particularly, $\nu_5$ and $\nu_6$) and scattered embryos~\citep{Raymond:2009is}. We discuss three classical models, in each the terrestrial disk extends from an inner edge to an outer edge carved by Jupiter. In the classical scenarios, the giant planets do not migrate, but they have significantly different orbits in each model due to different assumptions about past and future evolution of the Solar System. Each assumption has consequences for matching constraints such as the late heavy bombardment. Truncated disk models assume that either the terrestrial disk does not extend out to the giant planets or that the migration of the giant planets truncates the disk. This Grand Tack model is the most successful and is described in great detail.
\paragraph{Comparing terrestrial planet systems}
\begin{figure}[t!]
\begin{center}
\includegraphics[width=\textwidth]{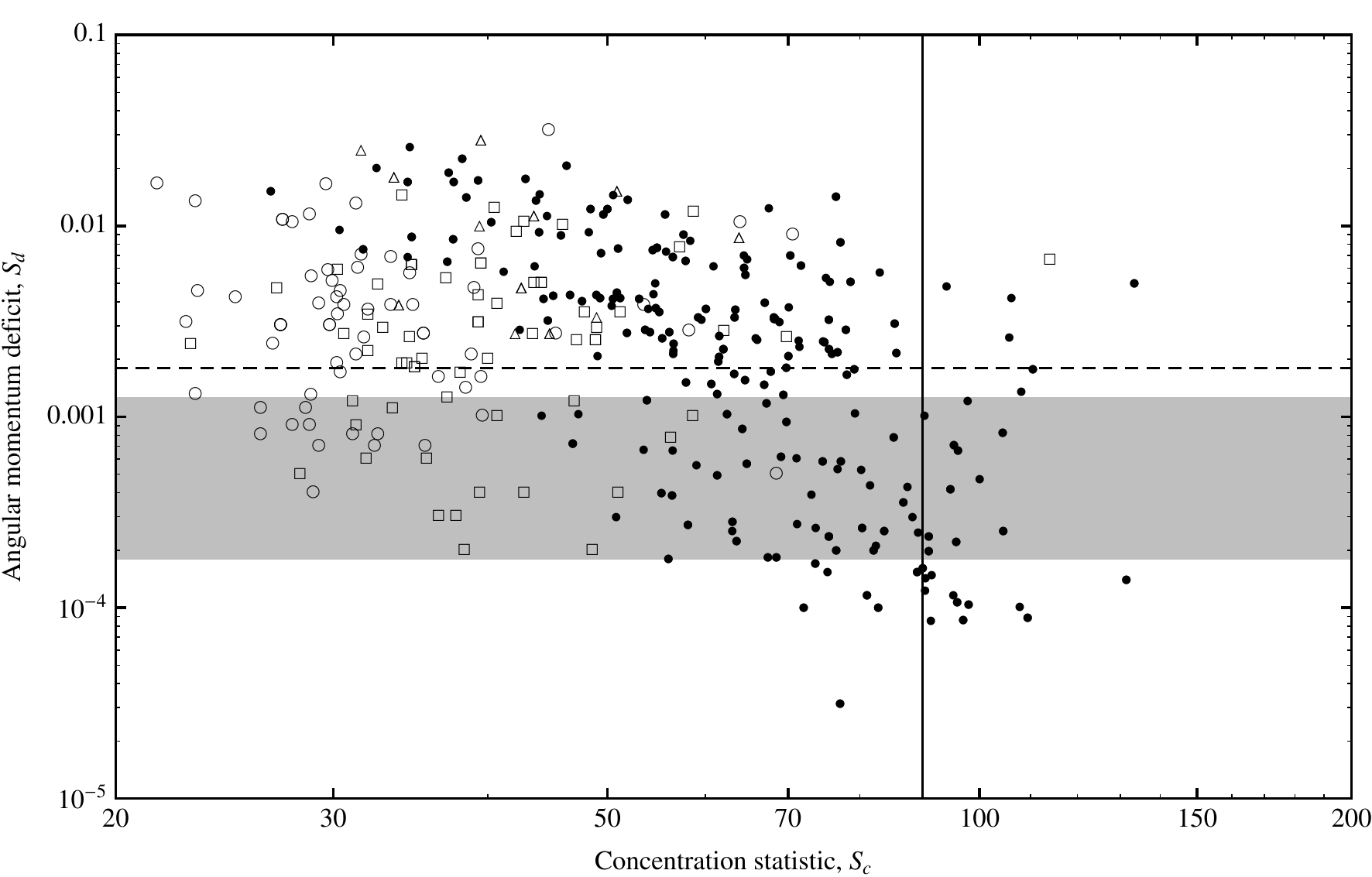}
\caption{The angular momentum deficit $S_d$ and concentration statistic $S_c$ for terrestrial planet systems formed from the eccentric~\citep[empty squares;][]{OBrien:2006jx,Raymond:2009is,Fischer:2014fm}, circular~\citep[empty circles;][]{OBrien:2006jx,Raymond:2009is,Fischer:2014fm}, and extra-eccentric~\citep[empty triangles;][]{Raymond:2009is} Jupiter and Saturn models and the Grand Tack model~\citep[solid points;][]{Walsh:2011co,OBrien:2014bk,Jacobson:2014cm,Jacobson:2014it}. The horizontal dashed line is the angular momentum deficit of the current terrestrial planet system $S_d = 0.0018$ and so the target value for the eccentric and extra-eccentric Jupiter and Saturn models. The gray rectangle marks the region with 10--70\% of the current angular momentum deficit~\citep{Brasser:2013hs}, and so is the target for models like the Grand Tack and circular Jupiter and Saturn that include a giant planet instability. The vertical solid line indicates the current terrestrial planet system's concentration statistic $S_c = 89.9$.}
\label{fig:scsdplot}
\end{center}
\end{figure}
Numerical simulations never exactly reproduce the terrestrial planets and cross-comparing the simulated planets with the true distribution of masses and orbits one-by-one is a complicated task. While, we can directly compare the orbits and masses of the planets in each system as shown in Figure 1 evaluating which model best reproduces the Solar System is a matter of statistics and gross metrics. To help with this process,~\citet{Chambers:2001kt} started using the following sophisticated quantities.

The angular momentum deficit $S_d$ of the inner terrestrial planets measures their dynamical excitation~\citep{Laskar:1997vw}. It is defined as the difference between the sums of the angular momentum of the corresponding circular and in-plane orbits and the actual orbit normalized by the sum of the angular momentum of the corresponding circular and in-plane orbits~\citep{Chambers:2001kt}:
\begin{equation}
S_d = \frac{\sum_j m_j \sqrt{a_j} \left( 1 - \sqrt{1 - e_j^2} \cos i_j \right)}{\sum_j m_j \sqrt{a_j}}
\end{equation}
where $m_j$ is the mass, $a_j$ is the semi-major axis, $e_j$ is the eccentricity and $i_j$ is the inclination of the $j$th planet. The inner planets do strongly perturb one another exchanging angular momentum, but exchanges with the outer planets are limited so the quantity $S_d$ is conserved within a factor of two~\citep{Laskar:1997vw}. The current value for the Solar System is $S_d=$ 0.0018 and is marked by a dashed horizontal line in Figure 2. If the Solar System undergoes a giant planet instability as supposed in the Nice model, then the angular momentum deficit after terrestrial planet formation needs to be between 10\% (0.00018) and 70\% (0.00126) of the deficit after the instability~\citep[shown in Figure 2 as the gray rectangle;][]{Brasser:2013hs}.

The other useful but unusual quantity is the concentration statistic $S_c$. The terrestrial planetary system of the Solar System is interesting because its mass is almost entirely in Earth and Venus with only a little bit in Mercury and Mars. The concentration statistic attempts, albeit degenerately, to capture this mass-orbit distribution with a single number. It is defined~\citep{Chambers:2001kt}:
\begin{equation}
S_c = \max \left[ \frac{\sum_j m_j}{\sum_j m_j \left( \log_{10} \left( a / a_j \right) \right)^2} \right]
\end{equation}
where like before $m_j$ is the mass and $a_j$ is the semi-major axis of the $j$th planet. $S_c$ is the maximum of the bracketed quantity as $a$ is varied. The terrestrial planets in the Solar System have $S_c = 89.9$, and this value is marked by a solid vertical line in Figure 2.
\subsection{Classical models}
In the classical models, the terrestrial disk of planetesimals and embryos has an outer boundary set by Jupiter, since most orbits exterior to the inner 3:2 mean-motion resonance with Jupiter are unstable. However, the inner boundary is more mysterious, and there is no conclusive theory regarding why the Solar System doesn't have planets interior to Mercury. Existing theories include high collision velocities that fragment and grind away the planetesimal population before embryos can form~\citep{Chambers:2001kt}, gas driven migration of embryos into the Sun~\citep{Ida:2008de}, an inner edge to the gas disk, or a silicate evaporation front that doesn't allow dust or pebbles to exist interior to form planetesimals and then embryos. Most N-body simulations place the inner edge at 0.3--0.7 AU to avoid forming planets interior to Mercury. Forming a Mercury-sized planet in the correct orbit with a satisfactory explanation for its high metal content is an open problem.
\paragraph{Eccentric Jupiter and Saturn, current orbits}
\label{sec:ejs}
The first giant impact phase models considered disks that contained about half the mass in the planetesimal population and half in the embryo population, which is consistent with the mass ratio found when oligarchic disks go unstable~\citep{Kokubo:1998ka}. Due to dynamical friction, these bodies start on low eccentricity and low inclination orbits. Importantly, Jupiter and Saturn are assumed to have formed on their current orbits, and the N-body simulations begin after the gas has been removed from the disk. 

This model is remarkably successful, it provided confidence that these models in general are on the right path. Early numerical simulations that only included a limited number of embryos naturally formed about four planets near the correct semi-major axes of the terrestrial planets (see Figure 1), giant impacts similar to the Moon-forming impact, and formation timescale of tens of millions of years for Earth-like planets~\citep{Agnor:1999ha,Chambers:2001kt}. When enough planetesimals are included in the simulations, dynamical friction removes excess energy and angular momentum from the orbits of the growing planets transferring it to the planetesimal population and the simulations reproduce the low eccentricities and low inclinations of the terrestrial planets~\citep{OBrien:2006jx,Morishima:2008bx}. Some but not all terrestrial planet systems created by the eccentric Jupiter and Saturn model have angular momentum deficits consistent with the Solar System~\citep{OBrien:2006jx,Raymond:2009is,Fischer:2014fm}; since the eccentric Jupiter and Saturn model is inconsistent with a Nice model-like giant planet instability, the most consistent angular momentum deficits are those closest to that of the current terrestrial planets (the dashed line in Figure 2).

\begin{figure}[t!]
\begin{center}
\includegraphics[width=\textwidth]{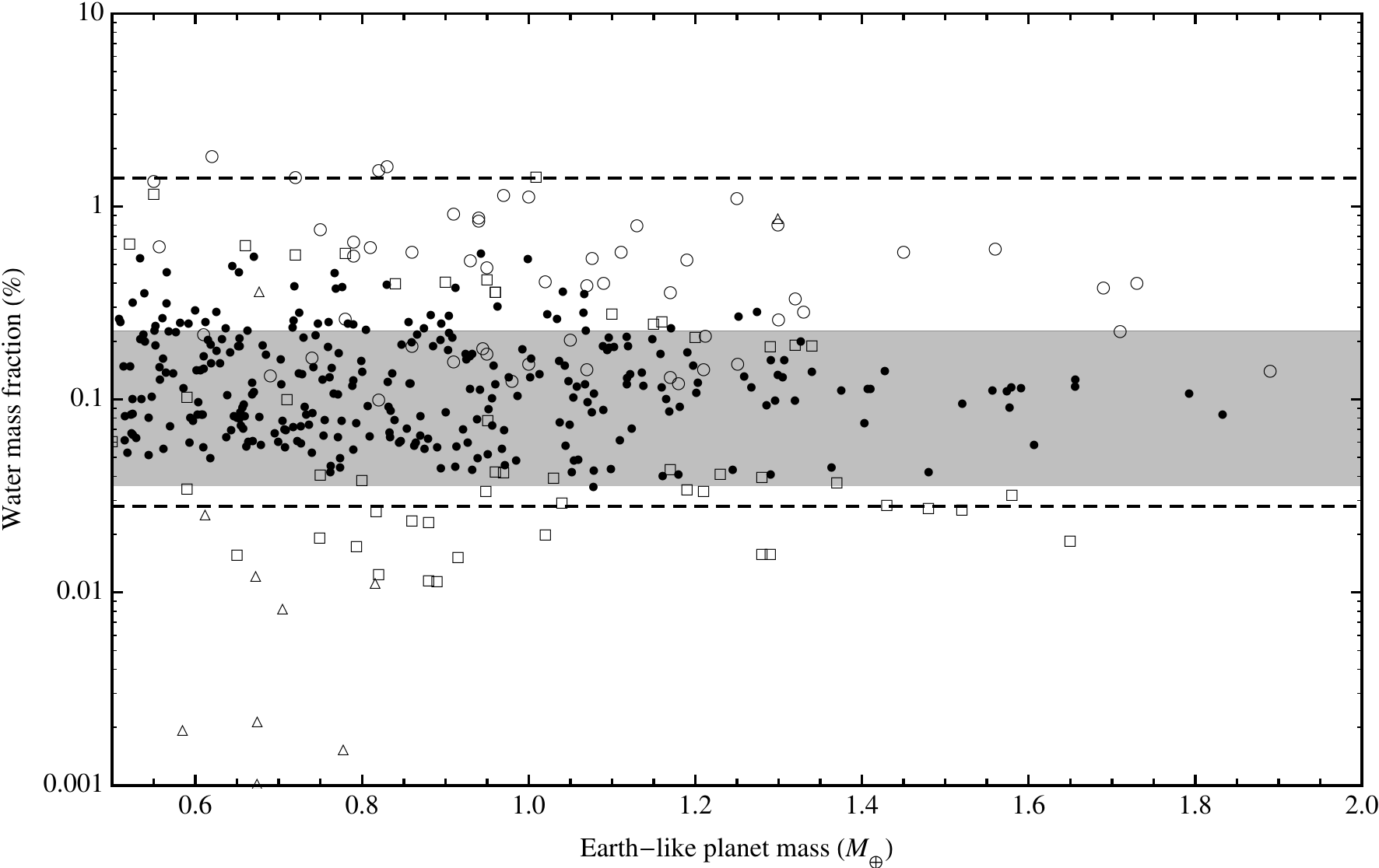}
\caption{The water mass fraction for each Earth-like planet formed from the eccentric~\citep[empty squares;][]{Raymond:2009is,Fischer:2014fm}, circular~\citep[empty circles;][]{Raymond:2009is,Fischer:2014fm}, and extra-eccentric~\citep[empty triangles;][]{Raymond:2009is} Jupiter and Saturn models and the Grand Tack model~\citep[solid points;][]{Walsh:2011co,OBrien:2014bk,Jacobson:2014cm,Jacobson:2014it} as a function of the mass of the planet. The water mass fraction is the water mass delivered to the planet assuming perfect accretion and either the water model of~\citet{Raymond:2004de} for the non-Grand Tack simulations or the model of~\citet{OBrien:2014bk} for the Grand Tack simulations. Earth-like planets from the numerical simulations have masses between 0.5 and 2 M$_\oplus$ and orbits between the current orbits of Mercury and Mars as shown in Figure~\ref{fig:massplot}. The upper dashed line is a liberal estimate of Earth's water content~\citep{Marty:2012gi}, the gray region is a more probable estimate~\citep{Lecuyer:1998wf}, and the lower dashed line is a minimum~\citep{Lecuyer:1998wf}, i.e. an estimate of the known water reservoirs. The water content of Earth's mantle reservoir is unknown.}
\label{fig:WMFplot}
\end{center}
\end{figure}
This model is haphazardly successful at delivering water to Earth-like planets from the outer asteroid belt via the $\nu_6$ secular resonance~\citep{Morbidelli:2000ex,Raymond:2004de,OBrien:2006jx,Raymond:2009is}. If most of Earth's water is delivered from the planetesimal carbonaceous chondrite parent bodies, then the D/H ratio of Earth's water is naturally explained~\citep[see][for greater detail and a discussion of the less likely alternatives]{Morbidelli:2000ex}. Inspired by the data from primitive meteorites, a simple model for the water content of planetesimals as a function of radius is that the water mass fraction is 10$^{-3}$\% interior of 2 AU, 10$^{-1}$\% between 2 and 2.5 AU, and 5\% exterior to 2.5 AU~\citep{Raymond:2004de}. This model is used for all of the classical scenario simulations in Figure 3, but not for the Grand Tack simulations since that model assumes that the carbonaceous chondrite parent bodies originate from exterior to Jupiter and Saturn.

\begin{figure}[t!]
\begin{center}
\includegraphics[width=0.98\textwidth]{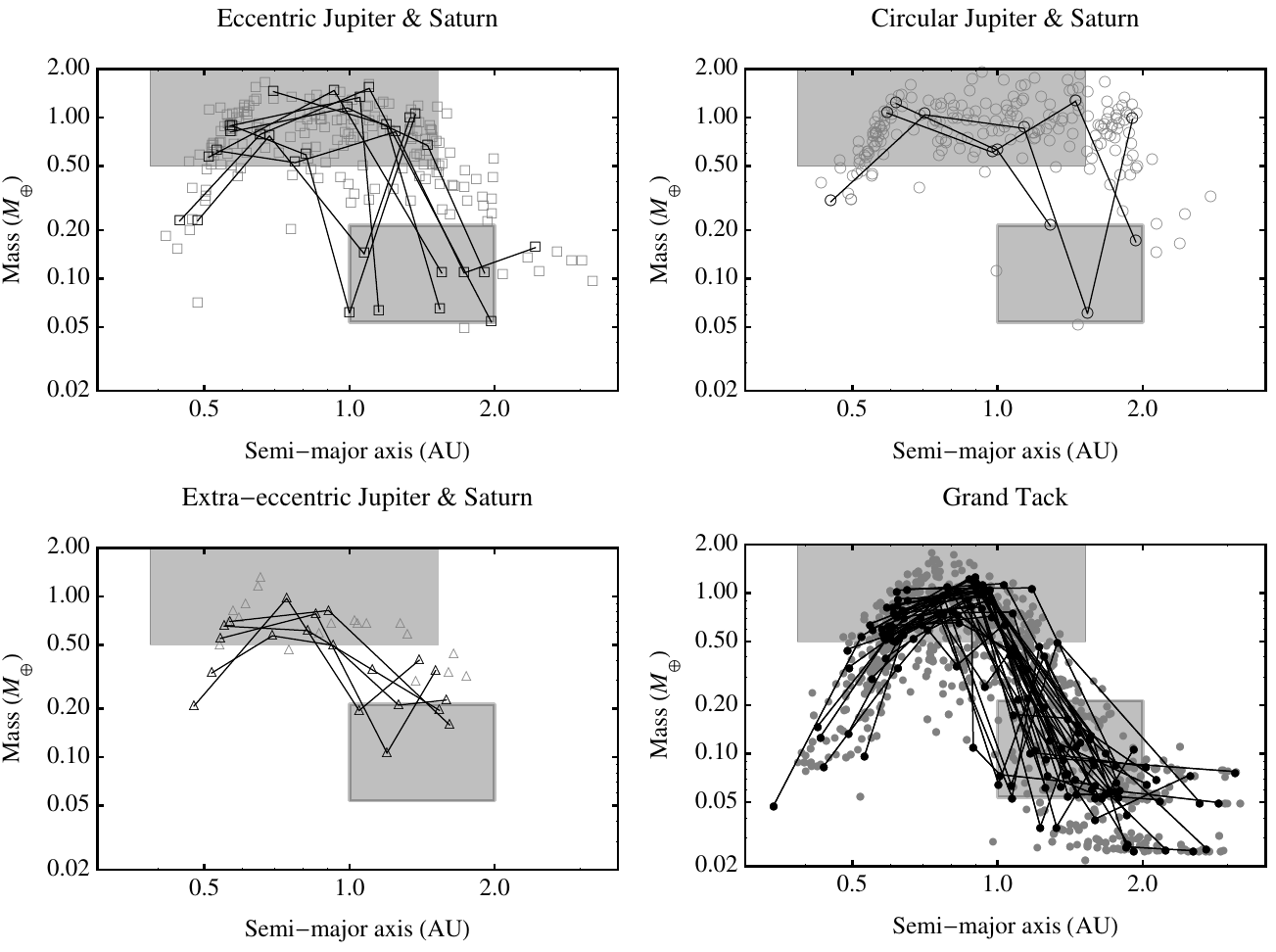}
\caption{Lines connect the final planets of the good terrestrial system analogs. Otherwise the individual graphs are very similar to Figure~\ref{fig:massplot}. For the classical models, good terrestrial system analogs have at least one Mars-like planet. For the Grand Tack model, good terrestrial system analogs have at least one Mars-like planet and both a Venus-like planet and Earth-like planet with no planets in between.}
\label{fig:MarsCompPlot}
\end{center}
\end{figure}
There are four outstanding issues with the eccentric Jupiter and Saturn model. First, this model systematically creates a planet at the position of Mars with a mass between 2 and 12 times that of Mars. As shown in Figure 4, there are a few exceptions to this rule and this directly leads to low values of the concentration statistic as shown in Figure 2. These were discovered by simulating a much larger number of simulations than had historically been studied before, emphasizing the need for this approach~\citep{Fischer:2014fm}. Of the 62 numerical simulations, 6 ($\sim$10\%) have Mars-like planets exterior of all Earth-like planets in the system. If Mars is an outlier of the eccentric Jupiter and Saturn model, then this is a challenge to the Copernican principle---that the Solar System is a usual rather than special outcome of planet formation processes. Before such a conclusion should be drawn, a more thorough investigation into what happened differently in these 6 simulations needs to be completed. 

\citet{Raymond:2009is} demonstrated that imperfect accretion via high velocity impacts cannot be responsible for the difference between the masses of the planets formed at Mars location and the mass of Mars. This was confirmed by~\citet{Chambers:2013cp}, who found that while imperfect accretion makes smaller planets overall including at Mars' location, it tends to decrease the mass ratio between Earth and Mars rather than increasing it relative to the perfect accretion simulations. Also, unlike Mercury, Mars appears inconsistent with this type of impact-driven mass loss.

Second, it does not explain how Jupiter and Saturn obtained their current eccentricities. Interactions with the nebular gas disk should leave the gas giants on quasi-circular orbits~\citep{Kley:2000us,Bryden:2000fn,Snellgrove:2001bu,Papaloizou:2003te,Crida:2008gp}. Also, giant planet migration is likely to move Jupiter and Saturn into a mean motion resonant configuration, and not into the near 5:2 mean motion configuration they currently occupy~\citep{Papaloizou:2001fb,Kley:2006er,Bitsch:2013cv}.

Third, this model is inconsistent with the Nice model, and so does not directly account for the late heavy bombardment. As discussed in Section~\ref{sec:Nice}, alternative explanations for the late heavy bombardment exist although none are as complete or successful as the Nice model. Only if an alternative is accepted, can the eccentric Jupiter and Saturn model be consistent with the evidence for the late heavy bombardment. 

Fourth, the eccentric Jupiter and Saturn model is internally inconsistent since interactions with planetesimals and embryos in the terrestrial disk tend to circularize the giant planets~\citep{Chambers:2001kt}.
\paragraph{Extra-eccentric Jupiter and Saturn}
\label{sec:eejs}
The extra-eccentric Jupiter and Saturn model was devised by~\citet{Chambers:2002hj} to alleviate the fourth outstanding issue of the previous model. The giant planet are assumed to have formed at their current semi-major axes but are on more excited orbits. This dynamical excitation is transferred through both scattering events and secular resonances, principally the $\nu_5$ and the $\nu_6$, to the terrestrial disk, and this creates angular momentum deficits much higher than those possessed by current terrestrial planets in the Solar System~\citep{Raymond:2009is}, as shown in Figure 2. 

The rapid depletion of the outer asteroid belt typically leads to a very dry Earth inconsistent with current water fraction estimates as shown in Figure 3. If the gas disk disperses such that Mars mass embryos in the outer asteroid belt region, assumed to be water-rich, are swept up in evolving secular resonances with Jupiter, then they may be incorporated into Earth delivering the needed water~\citep{Thommes:2008ep,Raymond:2009is}. Alternatively, water may be delivered by mechanisms listed by~\citet{Raymond:2009is} such as water adsorption of small silicate grains~\citep{Muralidharan:2008cs}, comet impacts~\citep{Owen:1995dx}, or oxidation of a H-rich atmosphere~\citep{Ikoma:2006co}.

Sweeping secular resonances combined with tidal gas drag can effectively deplete the asteroid belt by driving the inward migration of planetary embryos~\citep{Nagasawa:2005ga} and reproducing the small mass of Mars~\citep{Thommes:2008ep}, but this requires a rather specifically tuned gas disk, particularly the gas removal rate~\citep{Thommes:2008ep}. Sweeping resonances due to changing precession rates as the gas is dispersed is not necessary, fixed frequency resonant and secular perturbations do clear out the asteroid belt and Mars-region occasionally creating a small Mars, as shown in Figure 4, however there are often planetary embryos remaining in the asteroid belt~\citep{Raymond:2009is}. Thus, the extra-eccentric Jupiter and Saturn scenario weakly alleviates the first outstanding issue identified above, but many final planets in the Mars region are still too large and since the excitation retards the growth of Earth-like planets, they tend to be too numerous and too small. This leads to unchanged concentration statistics as shown in Figure 2.

The extra-eccentric Jupiter and Saturn scenario accentuates the second problem identified with the current orbit scenario; the assumption of high initial eccentricities for the giant planets is inconsistent with models of their growth in a gas disk~\citep{Morbidelli:2007ew,Pierens:2011ft,Bitsch:2013cv}. Similar to the eccentric Jupiter and Saturn scenario, this model is not coherent with the Nice model. 

Finally,\citet{Raymond:2009is} identifies a number of other explanations for extra-eccentric giant planets such as an early giant planet instability leaving the giant planets on very eccentric orbits and the corresponding problems with these models.
\paragraph{Circular Jupiter and Saturn, pre-Nice 2.0 model}
\label{sec:cjs}
Numerical simulations consistent with the Nice model automatically solve all of the latter three outstanding issues of the eccentric Jupiter and Saturn model~\citep{Raymond:2009is}. Giant planets growing and migrating in a gas disk circularize and enter resonances, and it is natural for Saturn and Jupiter to enter into a 3:2 mean motion resonance near 5.4 AU and 7.3 AU~\citep{Morbidelli:2007hy}. This configuration of the giant planets is primed for a giant planet instability at $\sim$400 My due to interactions with a massive outer disk of which the Kuiper Belt is a remnant~\citep{Levison:2011gt}.

This model has many of the same successes as the eccentric scenario: correct number of planets including Earth-like planets near 1 AU as shown in Figure 1, giant impacts similar to the Moon-forming impact~\citep{Raymond:2009is}, and growth timescales of tens of millions of years~\citep{Raymond:2009is,Fischer:2014fm}. Since the giant planets are on circular orbits, excitation due to secular resonances is reduced~\citep{Raymond:2009is,Fischer:2014fm} but since the giant planet instability in the Nice model will dynamically excite those planets, the target angular momentum deficit value is 10--70\% lower than for the eccentric Jupiter and Saturn model~\citep{Brasser:2013hs}. Such systems are successfully created as shown in Figure 2. Another success is the delivery of water to Earth from the asteroid belt. Using the same model as before~\citep{Raymond:2004de}, enough water from the outer asteroid belt is always delivered to Earth as shown in Figure 3~\citep{Raymond:2009is,Fischer:2014fm}.

Notably though, the small Mars problem is accentuated, as shown in Figure 4. Planets at the location of Mars are almost all between 2 and 20 times too massive. Only 1 ($\sim$2\%) of 62 systems has a Mars-like planet without also containing a third near-Earth mass planet. This reveals itself also as very low concentration statistics, as shown in Figure 2.

While this model is consistent with the Nice model, it is only partially consistent with evolution in a gas disk. The circularization and migration of the giant planets into mean motion resonances in the circular Jupiter and Saturn model is assumed to have no effect on the evolution of the terrestrial disk. The Grand Tack model is the natural consequence of breaking this assumption.
\subsection{Truncated disk models}
As mentioned, one of the outstanding problems with the classical models of terrestrial planet formation is the size of the planet formed near 1.5 AU. As shown in Figure 1, these models systematically produce planets at the location of Mars that are about 5--10 times too massive.~\citet{Raymond:2009is} considered a broad range of models including those discussed above and a few others, but none consistently solved this problem and most introduced other issues as well. This problem is an old problem that has been recognized at least as far back as~\citet{Wetherill:1991wc}.~\citet{Walsh:2011je} considered planet migration and the resultant sweeping secular resonances as a way to deplete the region around 1.5 AU of solid material early enough to frustrate the accretion of Mars-analogs. This failed as the migration of the giant planets would happen too late compared to the accretion of Mars-like planets, leaving them at similar mass as in classical scenarios.

\citet{Hansen:2009ke} proposed a dramatic solution to this problem demanding wholesale changes to the surface density of the solid material in the disk. They found that if the initial disk is truncated at 1 AU then the mass distribution of the terrestrial planets could be recovered including the Earth/Mars mass ratio. The systematic success to the Hansen model stems not just from the disk truncation but the unstable dynamical over-packing of the 400 equal mass embryos, which begin the simulation with overlapping mutual Hill radii~\citep{OBrien:2014bk}. These initial conditions lead to the rapid scattering of some embryos out of the narrow annulus. The bodies scattered out beyond 1 AU were typically stabilized due to interactions with other scattered bodies and isolated beyond the edge of the disk. This allowed these bodies to stop their accretion early and remain at a low mass, resulting in a short accretion time relative to the planets that accrete within the annulus. This is similar physics to that observed in previous works that had disk edges around 1 or 1.5 AU due to computational requirements~\citep{Agnor:1999ha,Chambers:2001kt}.

\citet{Jin:2008ju} proposed that severe surface density depletions in the proto-planetary disk cause a truncation. Since different angular momentum transport mechanisms are active at different radii in the disk,~\citet{Jin:2008ju} found a non-monotonic gas surface density for the inner regions of the disk and a surface density minimum at about 1.5 AU.~\citet{Izidoro:2014cj} performed N-body experiments and deduced that to best fit Mars, a $\sim$~75\% surface depletion between 1.3 and 2 AU is needed, however the prediction by~\citet{Jin:2008ju} appear to be more consistent with $\sim$~25\% and over a narrower annulus. Unlike the hypothesis proposed in~\citet{Jin:2008ju},~\citet{Izidoro:2014cj} find that Mars is a stranded embryo scattered out from the interior region of higher surface density, similar to the Grand Tack. Dissimilar to the Grand Tack and like the classical scenario, the asteroid belt is grown from material already in the low surface density region in the~\citet{Jin:2008ju} and~\citet{Izidoro:2014cj} models.
\paragraph{`Grand Tack', migrating Jupiter and Saturn}
\label{sec:grandtack}
A narrow over-packed disk with a truncation of solid material at 1 AU produces ideal outcomes for the Earth/Mars mass ratio~\citep{Hansen:2009ke}, but why such a disk existed needs an explanation. However, a powerful means to re-distribute solid material in the early Solar System are the giant planets. The giant planets must have had a relatively short accretion timescale compared to that of the terrestrial planets (see Section~\ref{sec:history}) and so existed to move around terrestrial building block material. Hydrodynamic models and data from extra-solar planetary systems suggests that planets the size of Jupiter can carve an annular gap in a gaseous nebula resulting in their inward migration due to the viscous evolution of the disk~\citep{Ward:1980te,Lin:1986ef}. This migration, type-II migration, can lead to the inward migration of a Jupiter-mass planet on 100,000 year timescales. 

A series of works found that the presence of a second massive planet, a Saturn mass planet, can halt and even reverse the inward migration of a Jupiter mass planet~\citep{Masset:2001dk,Pierens:2008jg,Pierens:2014iw,Crida:2007kx,Pierens:2011ft,DAngelo:2012bu}. Saturn is found to migrate very close to Jupiter, reaching an exterior 2:3 mean motion resonance with Jupiter, and this proximity causes their gaps to overlap. The net effect is that Saturn's presence allows gas to pass through Jupiter's gap, stopping the inward type II migration. The lower surface density of gas behind Jupiter, owing to the presence of Saturn's gap, then allows for the inner disk to provide a much stronger torque on Jupiter's orbit pushing both planets outward.

The net effect of this inward-then-outward migration scenario is that it provides a possible mechanism to create a truncated disk of solids.~\citet{Walsh:2011co} found that if Jupiter reversed its inward migration at 1.5 AU, i.e. `tacked', then the disk of solids would be truncated at 1 AU, however the violent mechanism causing the truncation and over-packing the disk leaves a more excited initial state than that previously explored by~\citet{Hansen:2009ke}. Similarly, while~\citet{Hansen:2009ke} started with 400 similar-size bodies in the annulus, the~\citet{Walsh:2011co} model, dubbed `the Grand Tack', started with a bi-modal size distribution of embryos and planetesimals. Despite these differing initial conditions, the terrestrial planets produced were similar including the high Earth/Mars mass ratio as shown in Figure 1.

The Grand Tack creates an asteroid belt very differently than the classical models since Jupiter migrates through the asteroid belt--twice. Rather than depleting a pre-existing belt via secular resonances and scattering events, the asteroid belt in the Grand Tack is created by the inward scattering of material by the giant planets as they migrate outwards, but not all this material was originally from the outer Solar System. Some of it had been scattered into the outer Solar System from the inner Solar System originally.~\citet{Walsh:2011co} hypothesize that material from the outer Solar System corresponds to the broad C-complex of asteroids associated with the carbonaceous chondrites and the material that originated in the inner Solar System but was scattered out and then back in corresponds to the broad S-complex of asteroids associated with the ordinary chondrites. The classical models explain the observed, albeit messy, gradient from S-complex to C-complex in the asteroid belt~\citep{DeMeo:2013hw,DeMeo:2014hk} as a local property of the disk. However, the Grand Tack asserts that this transition occurs over a much larger radial section of the disk (interior and exterior to the giant planets), but the migration of the giant planets naturally re-creates this gradient in the asteroid belt~\citep{Walsh:2011co,Walsh:2012gy}.

This hypothesis has strong implications for delivery of water to Earth since the D/H ratios of carbonaceous chondrites best match Earth's water~\citep{Morbidelli:2000ex}. These outer planetesimals account for a total of 1--3\% of the total accreted mass for each of the terrestrial planets~\citep{Walsh:2011co,OBrien:2014bk}. As shown in Figure 3, they deliver enough water to Earth assuming that they contain $\sim$10\% water by mass. Moreover, unlike the classical models, the mass of C-complex material delivered to the planets is independently calibrated because the initial ratio of C-complex to S-complex material in the protoplanetary disk is known from the ratio that survives in the asteroid belt~\citep{DeMeo:2013hw,DeMeo:2014hk}.

The Grand Tack solves each of the outstanding problems identified with the eccentric Jupiter and Saturn model. The giant planets and the terrestrial disk evolve self-consistently in a gas disk and the Nice model is naturally incorporated. Venus-like, Earth-like and Mars-like planets are all natural outcomes and many systems have an architecture similar to the Solar System as shown in Figure 4. This is why the Grand Tack alone matches both the concentration statistic and the necessary angular momentum deficit, pre-Nice model in this case, as shown in Figure 2. 

There are still some open issues regarding the Grand Tack such as the formation of Mercury, which may just be a result of numerical resolution, and some discussed in~\citet{Raymond:2014tv}.
\section{Earth in the Grand Tack model}
\label{sec:growth}
\begin{figure}[t!]
\begin{center}
\includegraphics[width=\textwidth]{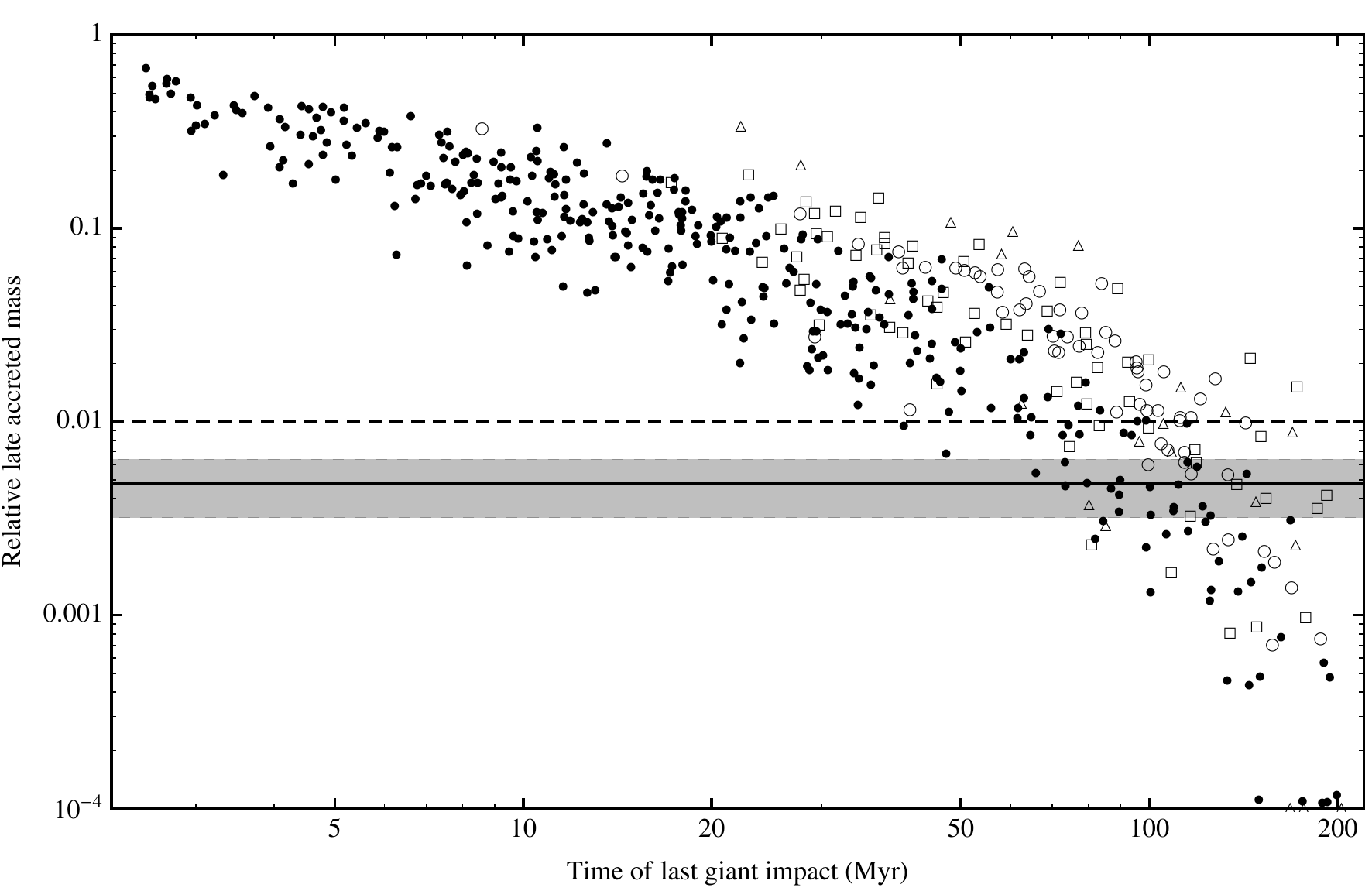}
\caption{The relative late accreted mass for each Earth-like planet formed from the eccentric~\citep[empty squares;][]{OBrien:2006jx,Raymond:2009is,Fischer:2014fm}, circular~\citep[empty circles;][]{OBrien:2006jx,Raymond:2009is,Fischer:2014fm}, and extra-eccentric~\citep[empty triangles;][]{Raymond:2009is} Jupiter and Saturn models and the Grand Tack model~\citep[solid points;][]{Walsh:2011co,OBrien:2014bk,Jacobson:2014cm,Jacobson:2014it} as a function of the time of the last giant impact. The relative late accreted mass is the mass accreted after the last giant (embryo) impact divided by the final mass of the planet. Earth-like planets from the numerical simulations have masses between 0.5 and 2 M$_\oplus$ and orbits between the current orbits of Mercury and Mars as shown in Figure~\ref{fig:massplot}.}
\label{fig:correplot}
\end{center}
\end{figure}
The Grand Tack model is best able to reproduce the mass-orbit distribution of the terrestrial planets and is the most consistent with the history of the Solar System. This includes reproducing the rapid growth of Mars~\citep{Dauphas:2011ec,Nimmo:2007bg} by acknowledging that Mars is a stranded embryo~\citep{Jacobson:2014it}. If Mars is an embryo then we have some information regarding the size of planetary embryos, and if planetesimals are born big, $\sim$100--1000 km, as the new story regarding pebble formation suggests, then the only remaining significant free parameter describing the bi-modal mass distribution is the ratio of the total mass of the embryo population to the total mass of the planetesimal population. This ratio has direct consequences for the dynamical friction present in the terrestrial disk: more planetesimal mass means more dynamical friction~\citep{Ida:1993ch}.

\citet{Jacobson:2014it} showed that increased dynamical friction also leads to an earlier last giant impact since lower relative velocities accelerate growth. Thus, the later the last giant impact, which is also the Moon-forming impact, then the embryo population must be more massive relative to the planetesimal population.~\citet{Jacobson:2014cm} showed there is a strong correlation between the time of the last giant impact and the late accreted mass, which is the mass accreted by Earth-like planets after the last giant impact. This makes sense because the mass of planetesimals in the disk decays with time, so the amount of planetesimal mass accreted after a later time is smaller than after an earlier time. This correlation for both the Grand Tack and the classical models is shown in Figure 5. Classical models always date the Moon-forming impact later than the Grand Tack.

The geologic record in Earth's mantle provides an estimate of the late accreted mass. From the highly siderophile element record on Earth~\citep{Becker:2006bi,Walker:2009be}, it appears that $\sim$5$\times$10$^{-3}$ Earth masses, M$_\oplus$, of material was delivered to Earth after the Moon-forming event~\citep[many lines of evidence restrict this late accreted mass to below 0.01 M$_\oplus$;][also see Morbidelli and Wood in this monograph]{Jacobson:2014cm}. Using the correlation between the late accreted mass and the time of the last giant impact, this late accreted mass estimate dates the Moon-forming impact to $\sim$95 My~\citep{Jacobson:2014cm}. Such a late last giant impact requires that most of the mass in the terrestrial disk be in the embryo population rather than the planetesimal population when Jupiter's migration interrupts the oligarchic growth phase~\citep{Jacobson:2014it}. This is consistent with models of the oligarchic growth phase that include planetesimal grinding~\citep{Levison:2012wp} as well as a pebble model that rarely creates planetesimals via the streaming instability but efficiently turns planetesimals into embryos via pebble accretion.
\subsection{The growth of Earth}
\label{sec:growth}
\begin{figure}[t]
\begin{center}
\includegraphics[width=\textwidth]{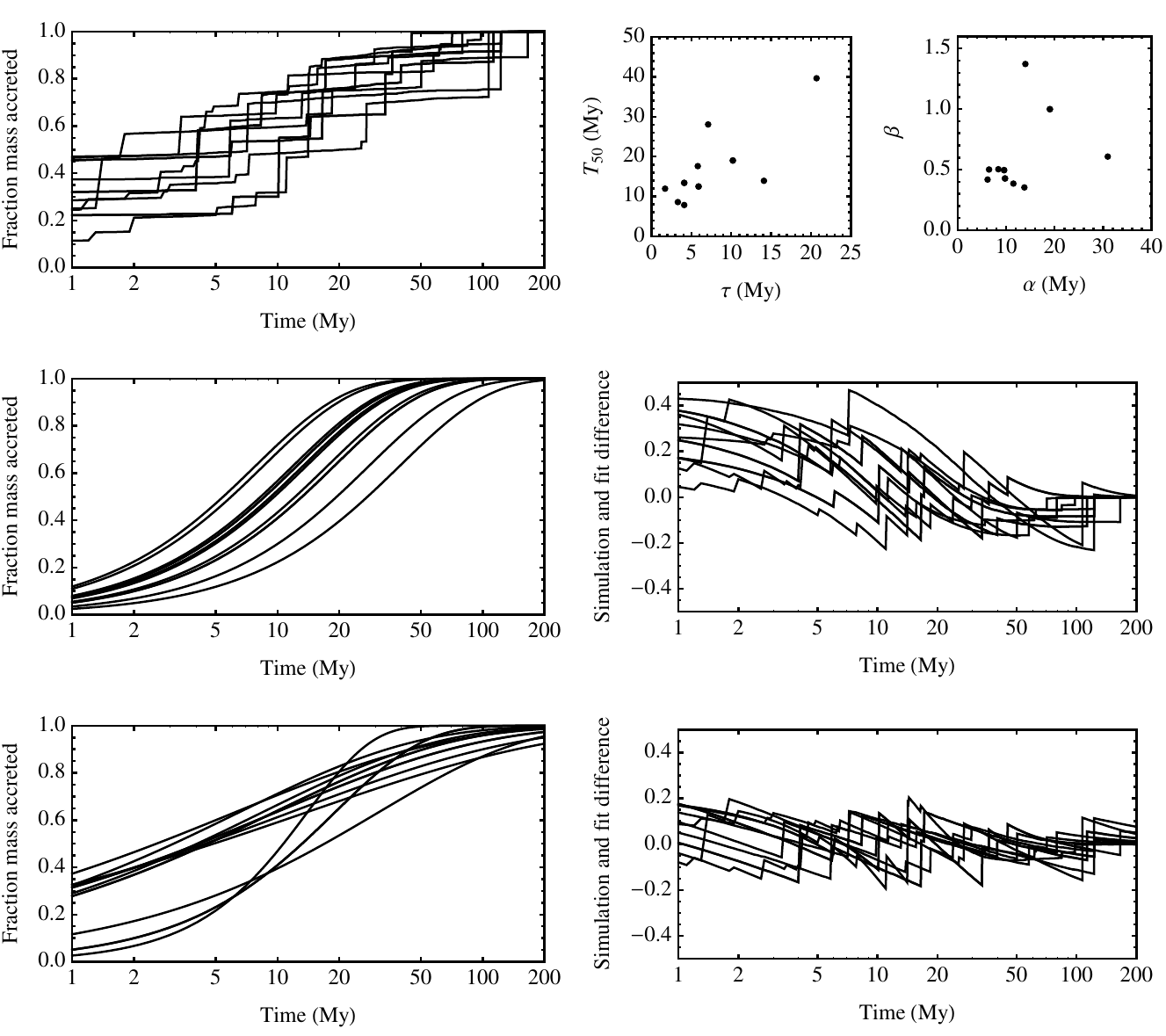}
\captionof{figure}[Figure]{Growth curve analysis for Earth-like planets. Upper left graph shows the growth of Earth-like planets as described in Section~\ref{sec:growth}, and the left middle and bottom graphs are fits using exponential and Weibull models, respectively. The upper center plot shows the time to reach 50\% of the final mass as a function of the exponential fit timescale and the upper right plot shows the $\alpha$ and $\beta$ of each of the Weibull fits. The right middle and bottom graphs are the the difference between the growth curve and the exponential and Weibull fits, respectively.}
\label{fig:GrowthCurvePlot}
\end{center}
\end{figure}
We now delve deep into the details of the growth of Earth in the Grand Tack scenario. In the prior sections, we were exploring the overall differences between the terrestrial planet formation models. Here, we focus on specific simulations that produce terrestrial planet systems most like the Solar System. In order to do that we make stricter requirements. We require that a Venus-like planet with mass between 0.4075 and 1.63 M$_\oplus$ be directly interior to an Earth-like planet with a mass between 0.5 and 2 M$_\oplus$ with at least one Mars-like planet exterior to both of those planets with a mass between 0.0535 and 0.114 M$_\oplus$. Furthermore, Earth-like planet most have a late accreted mass between 0.001 and 0.01 M$_\oplus$. These stringent requirements leave only 10 planetary systems out of the 203 Grand Tack numerical simulations from~\citet{Walsh:2011co},~\citet{OBrien:2014bk},~\citet{Jacobson:2014cm}, and~\citet{Jacobson:2014it} and all are from the latter two publications since they include the only simulations with a high enough total embryo to planetesimal mass ratio. Three of the ten have two Mars-like planets, but we include them for better statistics.

The growth of these Earth-like planets are shown in Figure 6. They all follow a very similar trajectory and grow quickly at first reaching 50\% of their growth in $T_{50}$$\sim$10 My, however their growth is strongly non-linear. In the past, many have approximated this non-linear growth with an exponential function $\propto \exp \left( t / \tau \right)$. The best fits to these simulations have exponential timescales $\tau$$\sim$3--10 My, but we find exponential fits to be inappropriate for the Grand Tack. Figure 6 shows exponential fits to the growth profile of each Earth-like planet, and then shows the difference between the actual growth trajectory and those fits. There is a clear underestimate of the growth rate at early times. A much better model is the Weibull model $\propto \exp \left( t / \alpha \right)^{\beta}$. The best fits to these simulations are shown in Figure 6 with $\alpha$$\sim$10 My and $\beta$$\sim$0.5. This model has no systematic over- or under-estimate. However, even this model can be off by 20\%.
\subsection{Composition of Earth and the other terrestrial planets}
\begin{figure}[t!]
\begin{center}
\includegraphics[width=\textwidth]{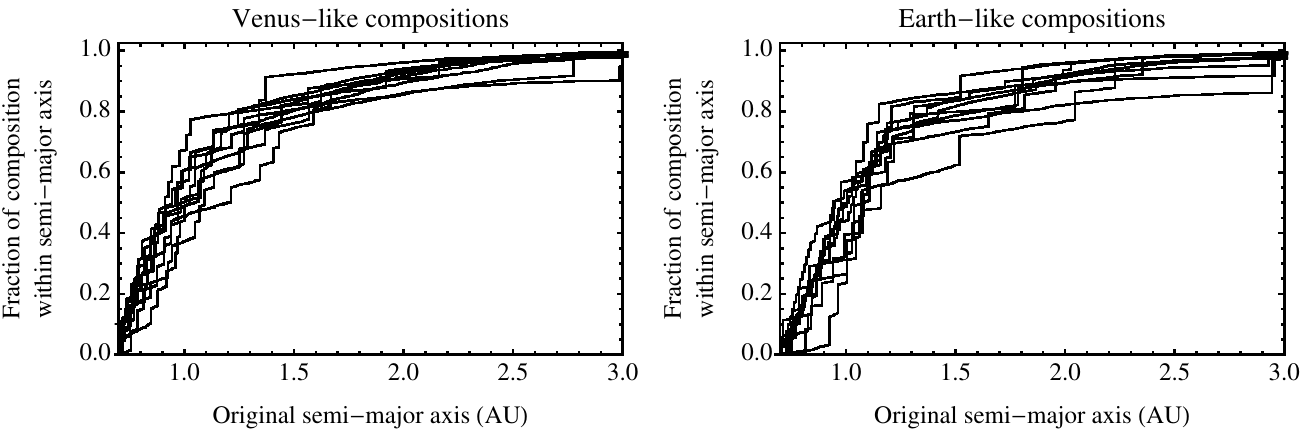}
\caption{The composition curves for Venus-like and Earth-like planets. These curves correspond to the terrestrial planetary systems described in Section~\ref{sec:growth}. The composition curves show the cumulative mass in the final planet as a function of semi-major axis. Some mass is accreted from exterior of 3 AU.}
\label{fig:EarthVenusComp}
\end{center}
\end{figure}
In the Grand Tack, the inner Solar System is thoroughly radially mixed by the inward migration of Jupiter~\citep{OBrien:2014bk}. This naturally produces Venus-like and Earth-like planets that are compositionally very similar. Both planets contain multiple embryos and many planetesimals from throughout the disk. Figure 7 shows their compositions as a cumulative function of semi-major axis. There are three different compositional regions made clear. First, $\sim$75\% of the mass originates from $0.7$ AU, the inner edge of the terrestrial disk, to $\sim$1.1 AU, the location of the 3:2 inner mean motion resonance with the location of Jupiter's `tack' at 1.5 AU. The mass from within this region is remarkably linearly representative. About $\sim$22--25\% of each planet's mass originates from exterior of 1.1 AU to the outer edge of the terrestrial disk at about $4$ AU. This mass is not delivered linearly with semi-major axis, but with slightly more mass delivered from the inner section than the outer. This is because material scattered by Jupiter further out in the disk was less likely to be scattered all the way into the 1 AU region where the proto-Earth and proto-Venus were growing. The final $\sim$1--3\% of mass comes from the outer disk exterior to the giant planets~\citep{OBrien:2014bk}.

\begin{figure}[t!]
\begin{center}
\includegraphics[width=\textwidth]{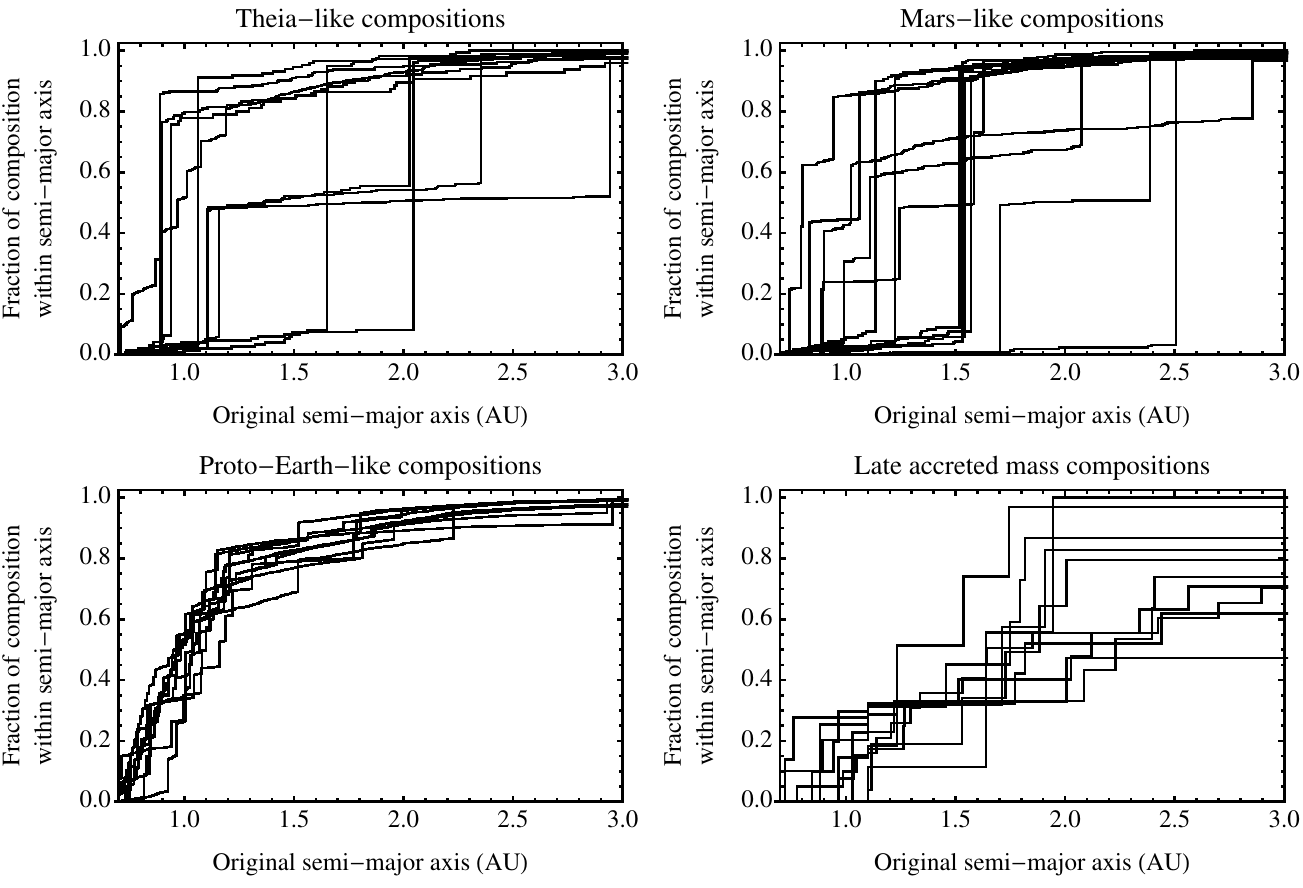}
\caption{The composition curves for the Theia-like, Mars-like, and proto-Earth-like planets and the late accreted mass composition. These curves correspond to the terrestrial planetary systems described in Section~\ref{sec:growth}. The composition curves show the cumulative mass in the final planet as a function of semi-major axis. Some mass is accreted from exterior of 3 AU.}
\label{fig:OtherComp}
\end{center}
\end{figure}
The feeding zones of the proto-Earth before the last giant (Moon-forming) impact is shown for each simulation in Figure 8. It very much resembles the final Earth-like planet, which makes sense because Theia, the last giant impactor, and the late accreted mass, which are also shown in Figure 8, contribute only a small amount of mass. The late accreted mass appears to sample the entire inner disk. Unlike the final Earth-like planet, the composition of the late accreted mass doesn't reflect the location of Jupiter's tack. 

Theia is usually a stranded embryo like Mars, although sometimes Theia is much larger and looks very similar to the proto-Earth. The feeding zones of Mars-like planets are also shown in Figure 8 and these reveal that while Mars-like planets often come from the outer disk, they do not always. Similarly, while Theia often comes from the inner terrestrial disk near 1 AU, it can come from the outer terrestrial disk. Some Mars-like planet compositions are composed of different embryos and so are unlikely to match the Hf-W growth constraints.

\section{Conclusion and discussion}
The growth of Earth cannot be isolated from the formation of the rest of the Solar System. While the number of constraints regarding the history of the Solar System are few, they are powerful. The mass and orbit distribution of the terrestrial planets and the evidence for the late heavy bombardment rule out the classical terrestrial plant formation models. Trunfcated disk models reproduce the mass and orbit distribution of the terrestrial planets particularly the Earth/Mars mass ratio. Of these models, only the Grand Tack is completely self-consistent and consistent with a late heavy bombardment via the Nice model.

The growth of Earth in the Grand Tack model follows a Weibull accretion curve to within 20\%. During its growth, it samples $\sim$75\% of its mass linearly from the inner terrestrial disk edge to the location of the 3:2 inner mean motion resonance with the location of Jupiter's tack. The remainder of its mass comes predominately from the rest of the terrestrial disk but biased towards the regions exterior but closes to 1.1 AU. Plenty of water is delivered to Earth via C-complex asteroids from the outer Solar System. The total mass of these asteroids delivered into the inner Solar System is calibrated by the ratio of S-complex to C-complex asteroids in the Main Belt. 

From the highly siderophile record on Earth, the age of the Moon is estimated to be $\sim$95 My~\citep{Jacobson:2014cm}. Since this age estimate is high, the ratio of the total mass in the embryo population to the total mass in the planetesimal population must also be high~\citep{Jacobson:2014it}. Since Mars is a stranded embryo~\citep{Nimmo:2007bg,Dauphas:2011ec}, the mass of individual embryos is also known~\citep{Jacobson:2014it}. These two facts strongly constrain the structure of the terrestrial disk at the time of Jupiter's inward migration. While this could be explained by planetesimal grinding, it could also be the result of pebble processes.

Pebble processes are likely the key to growing past the boulder barrier and creating planetesimals via the streaming instability. They also likely contribute a significant amount of mass to embryos via pebble accretion. These processes are likely to leave unique compositional gradients and signatures in the disk. Combining these gradients with models of the composition of Earth and observations of Earth's composition will be a powerful test of these hypotheses.

\clearpage

\bibliography{/Users/seth/Papers/biblio}
\bibliographystyle{agufull08.bst}

\end{document}